\documentclass[prl,superscriptaddress,twocolumn,aps,showpacs]{revtex4}
\usepackage{amssymb}
\usepackage{graphicx}
\usepackage{longtable}
\usepackage{amsmath}
\usepackage{amsfonts}
\usepackage{color}
\begin{document}
\author{A.~Chiesa}
\affiliation{Dipartimento di Fisica e Scienze della Terra, University of Parma, 43124 Parma, Italy}
\affiliation{Institute for Advanced Simulation, Forschungszentrum J\"ulich,
             52425 J\"ulich, Germany}
\author{S.~Carretta}
\affiliation{Dipartimento di Fisica e Scienze della Terra, University of Parma, 43124 Parma, Italy}
\author{P.~Santini}
\affiliation{Dipartimento di Fisica e Scienze della Terra, University of Parma, 43124 Parma, Italy}
\author{G.~Amoretti}
\affiliation{Dipartimento di Fisica e Scienze della Terra, University of Parma, 43124 Parma, Italy}
\author{E.~Pavarini}
\affiliation{Institute for Advanced Simulation, Forschungszentrum J\"ulich,
             52425 J\"ulich, Germany}
\affiliation{JARA High-Performance Computing}
\date{\today }
\title{Many-body models for molecular nanomagnets}
\begin{abstract}
We present a flexible and effective {\em ab-initio} scheme to build many-body models for molecular nanomagnets,
and to calculate magnetic exchange couplings and zero-field splittings.
It is based on using localized Foster-Boys orbitals as one-electron basis.
  We apply this scheme to three paradigmatic systems, the antiferromagnetic rings Cr$_8$ and Cr$_7$Ni and the single molecule magnet Fe$_4$. In all cases we identify the essential magnetic interactions and find excellent agreement with experiments.
\end{abstract}
\maketitle

\begin{figure}
   \centering
   \includegraphics[height=3.2in]{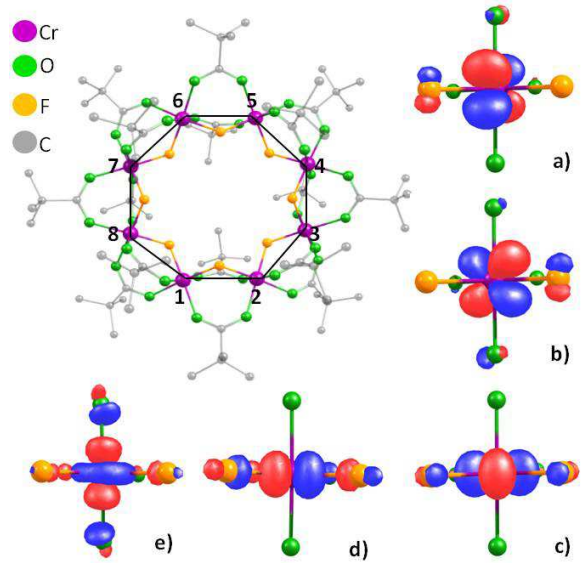} %
   \caption{{
   The Cr$_8$ antiferromagnetic ring and its Cr $d$ crystal-field orbitals for site 1, in order  $a)\to e)$ of increasing energy.
   Red (blue) orbital lobes  are positive (negative).
   The covalent $p$ tails on the neighboring ligands carry the information on the molecular structure and are crucial for the magnetic exchange couplings.
   We define $z$ as the axis perpendicular to the ring and pointing outwards. All Cr sites are approximatively equivalent. {H atoms are not shown for clarity}.
}}
   \label{Cr8}
\end{figure}

Clusters made of a finite number of interacting spins are ideal test beds to investigate fundamental issues in quantum mechanics.
One of the first physical realizations are molecular nanomagnets (MNMs),
molecules containing a core of $d$ or $f$ ions, whose spins are coupled by magnetic %
interactions; MNMs form crystals which behave like an ensemble of identical and almost non-interacting magnetic units. %
During the last years sophisticated experiments and targeted research activities  have unveiled a variety of fundamental quantum phenomena and potential technological applications of MNMs \cite{Gatteschibook,SessoliNature,SessoliNaturemat,Ni10,LossNature,QC1,NatureNano}.
The two most promising classes  have been identified in the Single-Molecule Magnets (SMM), like Mn$_{12}$ \cite{Gatteschibook}, Fe$_8$\cite{Gatteschibook} and Fe$_4$ \cite{Gatteschibook,Fe4INS,Fe4Chimica}, and the Antiferromagnetic Rings (AFR), like Cr$_8$ \cite{Cr8PRB,Cr8VanSL},
shown in Fig.~\ref{Cr8}, Cr$_7$Ni \cite{QC1,INSCr7Ni,BlundellCr7Ni2} and Fe$_6$ \cite{Gatteschibook}. While SMM have opened the perspective of storing information in single molecules and building high-density magnetic memories, AFR are of great interest in the field of quantum information processing
\cite{QC1,BlundellCr7Ni2,NatureNano,PRLent,PRLsim}.

At the synthetic level, thanks to the huge progresses made in the last years in coordination chemistry, it is now possible to reach a high degree of control on the molecular structure and on the topology of magnetic interactions.
At the theoretical level, one of the main obstacles to further progress remains the lack of a flexible and systematic approach to build {\em ab-initio} system-specific models for the magnetic interactions; such models should describe on the same footing chemistry and many-body effects within the partially filled $d$ or $f$ shells of the magnetic ions.

MNMs are typically described through Heisenberg-like spin Hamiltonians. If the form of the spin Hamiltonian is known,  
the magnetic couplings can be in principle extracted from total-energy density-functional theory (DFT) calculations for different spin configurations \cite{pedersondft, others,b3lyp,lda+u2}.
This approach can become unpractical if many parameters have to be determined, as e.g. in heterometallic compounds or anisotropic SMM; furthermore,
subtle interactions, which could greatly influence, e.g., the relaxation dynamics, can be easily overlooked. An alternative consists in computing
the couplings via energy variations at small spin rotations \cite{lda+u}.
However, at a more fundamental level, a common problem of all these approaches is that the most used DFT functionals (the local-density approximation (LDA) and its simple extensions), do not properly describe strong correlation effects in open $d$- or $f$-shells, while LDA+$U$  or hybrid functionals include them only at the static mean-field level. Recently, it has been suggested that Hubbard-like models could be more appropriate
 \cite{trif2010,pedersonhubbard},
but an efficient and flexible scheme to calculate the parameters of such models {\em ab-initio}, no matter the complexity of the system, has not been implemented so far.

\begin{figure}[t]
   \centering
   \includegraphics[height=3.0in]{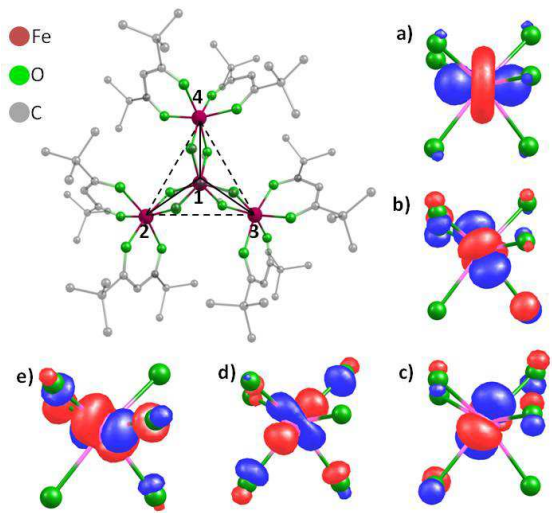} %
   \caption{{Structure of Fe$_4$  and Fe $d$ crystal-field orbitals localized on the central Fe ($D_3$ symmetry):
   The $a_1$ singlet $(a)$, and the two $e$ doublets $(b,c)$ and $(d,e)$, in order of increasing energy.   Heisenberg couplings: $\Gamma^{1,2}$ (full line) and $\Gamma^{2,3}$ (dashed line).}
   We define $z$ as the axis perpendicular to the triangle, and pointing outwards. {H atoms are not shown for clarity}.}
   \label{Fe4}
\end{figure}

In this Letter we show that this can be achieved by using localized Foster-Boys orbitals \cite{Boys} as one-electron basis to construct molecule-specific generalized Hubbard models. We
use the constrained local-density approximation (cLDA) scheme \cite{cLDA} to calculate the screened Coulomb interactions in such a Foster-Boys basis.
We obtain the spin Hamiltonians systematically by using a canonical transformation \cite{macdonald} to eliminate charge fluctuations, without any {\em a-priori} assumption on the form of the final spin Hamiltonian.
We implement this scheme in the NWChem quantum-chemistry code \cite{nwchem}, and apply it to three prototype molecules, representative of the two main classes of MNMs: the AFRs Cr$_8$  (Fig.~\ref{Cr8}) and Cr$_7$Ni and the SMM Fe$_4$ (Fig.~\ref{Fe4}).
These systems have been extensively investigated experimentally \cite{INSCr7Ni,Cr8PRB,Cr7Nitorque,Fe4INS,Fe4Chimica} and the magnetic exchange couplings are now well known.
In all cases, we find excellent agreement with experiments, and identify the microscopic mechanisms that lead to the empirical spin models commonly adopted to describe them.

The procedure we adopt is the following.
First we perform LDA calculations for the experimental structures reported in Refs.~\cite{molecularstructures};
in this step we use as basis a triple-zeta valence set of gaussians. Next, we identify the transition-metal $d$-like molecular orbitals; by means of Foster-Boys localization \cite{Boys},
we construct a set of localized orbitals, which are centered on the transition-metal ions and span such  
 $d$-like states;
using these orbitals as basis we build the corresponding generalized Hubbard model
\begin{eqnarray}\nonumber
H&=&-\sum_{ii^\prime\sigma}\sum_{mm^\prime} t^{i,i^\prime}_{m,m^\prime} c^\dagger_{im\sigma} c^\dagger_{i^\prime m^\prime\sigma} \\ \nonumber
&+&\frac{1}{2} \sum_{i i^\prime\sigma\sigma^\prime}\sum_{mm^\prime}\sum_{p  p^\prime}
U^{i,i^\prime}_{m p m^\prime p^\prime}
c^\dagger_{im\sigma} c^\dagger_{ip\sigma^\prime} c^{\phantom{\dagger}}_{i^\prime p^\prime\sigma^\prime}c^{\phantom{\dagger}}_{i^\prime m^\prime\sigma} \\ \label{hubbard}
&+&\sum_i \lambda_i \; {\bf S}_i\cdot {\bf L}_i -H_{\rm DC}.
\end{eqnarray}
Here $c_{im\sigma}^\dagger$ ($c_{im\sigma}^{\phantom{\dagger}}$) creates (annihilates) an electron with spin $\sigma$ in the Boys orbital $m$ at site $i$. The parameters
$t^{i,i^\prime}_{m,m^\prime}$ are the hopping integrals ($i\ne i^\prime$) or the crystal-field matrix ($i=i^\prime$), while
$U^{i,i^\prime}_{m p m^\prime p^\prime}$ are the screened Coulomb integrals (Tab.~\ref{hoppings}).
The term $H_{\rm DC}$ is the double counting correction,
which removes the part of the Coulomb interaction already included and well accounted for in the LDA; $\lambda_i$ is the spin-orbit coupling. %
\begin{table}[b]
   \centering
   \begin{tabular}{@{} lrr@{\;\;\;}c|rrrrrr @{}}
    \multicolumn{9}{c} {\phantom{Ni}Cr$_8$} \\ \hline
    \multicolumn{2}{c}{$\quad\varepsilon_n$}&\multicolumn{7}{c}{$t_{n,n^\prime}^{i,i+1}$} \\[1ex]
       $|n\rangle$& &&
       \begin{picture}(25,20)(0,0)
        	\put(13,10){$|n^\prime \rangle$}
          \put(0,19){\line(5,-4){28}}
          \put(5,0){$|n\rangle$}
     \end{picture}&
       \begin{picture}(20,10)(0,0)
        	\put(4,10){$|1 \rangle$}
     \end{picture}&
       \begin{picture}(20,10)(0,0)
        	\put(4,10){$|2 \rangle$}
     \end{picture}&
       \begin{picture}(20,10)(0,0)
        	\put(4,10){$|3 \rangle$}
     \end{picture}&
       \begin{picture}(20,10)(0,0)
        	\put(4,10){$|4 \rangle$}
     \end{picture}&
       \begin{picture}(20,10)(0,0)
        	\put(4,10){$|5 \rangle$}
     \end{picture}&
     \\
       \cline{1-2}\cline{4-9}
       $|1\rangle$   &-0.071  &&~ $|1\rangle$~~~~&-0.231&	0.041&	-0.001&	0.056&	0.028 \\
       $|2\rangle$   &-0.061  &&~ $|2\rangle$~~~~&-0.057&	0.085&	-0.061&	-0.019&	0.010 \\
       $|3\rangle$   & 0.040  &&~ $|3\rangle$~~~~&\phantom{-}0.011&	 \phantom{-}0.021&	0.033&	-0.154~&	-0.160 \\
       $|4\rangle$   &2.021   &&~ $|4\rangle$~~~~&-0.092&	-0.128&	-0.171&	0.094&	0.164 \\
       $|5\rangle$   &2.070   &&~ $|5\rangle$~~~~&\phantom{-}0.001	&-0.053	&-0.011&	0.114&	-0.033
   \end{tabular}\\[3ex]
   \centering
   \begin{tabular}{@{} l|rrrl @{\,}}
      &\phantom{Ni}{Cr$_8$}&\phantom{Ni}{Fe$_4$}& \multicolumn{2}{l}{~~{Cr$_7$Ni}} \\
      [0.5ex]
      \hline
      $U^{1,1}$      &5.98 &5.22& ~~6.32  &(Ni)\\
      $U^{2,2}$      &5.98 &5.03& ~~5.98  &(Cr)\\
      $J^{1,1}$     &0.26&0.24&~~0.21 &(Ni)\\
      $J^{2,2}$     &0.26&0.22&~~0.26 &(Cr)\\
      $\lambda_1$  &16.5&34.3&~~33.5&(Ni)\\
      $\lambda_2$  &16.5&37.0&~~16.5&(Cr)\\
    \end{tabular}
   \caption{
   \label{hoppings} Top: Crystal-field energy levels and hopping integrals for Cr$_8$. The latter are given in the basis of crystal-field orbitals (in eV), and for sites $i=1$ and $i^\prime=2$ in Fig.~\ref{Cr8}. The energy of the Fermi level is set to zero.
   Bottom: Screened Coulomb integrals $U^{i,i}$ and $J^{i,i}$ obtained via cLDA.  
   Sites $i$ are defined in Fig.~\ref{Cr8} and Fig.~\ref{Fe4}.
    }
\end{table}
The results presented in this work are obtained using
the rotational invariant form of the Coulomb vertex, including spin-flip and pair hopping terms
but (for simplicity) no Coulomb anisotropy;
thus all Coulomb parameters can be expressed as a function of the averaged screened Coulomb couplings $U^{i,i}$ and $J^{i,i}$ \cite{explanations}.
We determine the latter by using the cLDA \cite{cLDA} approach in the Foster-Boys basis, keeping the basis frozen in the self-consistency loop.
For $H_{\rm DC}$ we adopt the common \cite{book} expression $H_{\rm DC}=\frac{1}{2}\sum_iU^{i,i}n^i_d (n^i_d-1)-\frac{1}{4}\sum_iJ^{i,i} n^i_d(\frac{1}{2}n^i_d-1),$ where $n^i_d$ is the number of $d$ electrons at site $i$.
For homonuclear systems  $H_{\rm DC}$ amounts to a  shift of the $d$ levels and can be incorporated in the chemical potential; in the case of Cr$_7$Ni, instead, the shift due to $H_{\rm DC}$ has to be taken into account explicitly.
Finally, we extract the spin-orbit coupling $\lambda_i$ by comparing the one-electron part of Hamiltonian (\ref{hubbard}) obtained with and without spin-orbit interaction.
Once we have obtained the parameters of the Hubbard model (Tab.~\ref{hoppings}), by using a canonical transformation, we eliminate charge fluctuations and
derive the corresponding low-energy spin model. In this step, it is convenient to work in the basis
of crystal-field states, obtained by diagonalizing the on-site matrix $t^{i,i^\prime}_{m,m^\prime}$; we denote their energies $\varepsilon_n$ with $\varepsilon_n \le\varepsilon_{n+1}$.
The crystal-field states are shown in Fig.~\ref{Cr8} for Cr$_8$ and Fig.~\ref{Fe4} for Fe$_4$.
At all sites but the central Fe in Fe$_4$,
the environment of the magnetic ion is approximatively octahedral; thus
the crystal-field orbitals split into a lower energy $t_{2g}$-like {quasi-triplet}
and a $1-2$~eV higher energy $e_g$-like {quasi-doublet}.
The central Fe site of Fe$_4$ has D$_3$ symmetry; its crystal-field levels (Fig.~\ref{Fe4})
split into a $a_1$ ground state and two excited $e$ doublets, $\sim 0.6$~eV and
$\sim 1.7$~eV above.

%
\begin{figure}[t]
   \centering
   \includegraphics[height=2.5in]{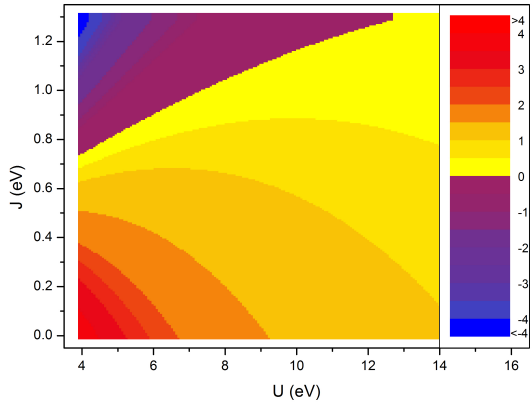} %
   \caption{Cr$_8$: Calculated super-exchange coupling (in meV) between nearest neighbors \cite{non-equiv}, $\Gamma_{\rm SE}^{i,i+1}$,   as a function of the Coulomb parameters $J$ and $U$. Calculations are performed using the rotationally-invariant Coulomb interaction.
   The cLDA values of the screened Coulomb integrals
   are $U\sim 6$~eV and $J\sim 0.26$~eV. Around these values  $\Gamma_{\rm SE}^{i,i+1}>0$ (AFM). 
    }
   \label{JvsJU}
\end{figure}

For all systems analyzed, we find that the essential spin interactions are described by the spin Hamiltonian
\begin{equation}
H\!=\! \frac{1}{2} \sum_{ii^\prime}\Gamma^{i,i^\prime} {\bf S}_i\cdot {\bf S}_{i^\prime}\! +\! \sum_i D^i \left[ S_{iz}^2-\frac{1}{3}S_i (S_i+1)\right],\label{HS}
\end{equation}
where $\Gamma^{i,i^\prime}$ are the isotropic magnetic couplings and  $D^i$ 
a zero-field splitting parameters (ZFS),
which is negative if $z$ is an easy axis; the $z$ direction is defined as the axis perpendicular to the ring
(Fig.~\ref{Cr8}) or to the triangle (Fig.~\ref{Fe4}).
The coupling $\Gamma^{i,i^\prime}=\Gamma^{i,i^\prime}_{\rm CE}+\Gamma^{i,i^\prime}_{\rm SE}$ is the sum of a ferromagnetic (FM) screened Coulomb exchange term, $\Gamma^{i,i^\prime}_{\rm CE}$,  which we obtain via cLDA calculations, and
a super-exchange term $\Gamma^{i,i^\prime}_{\rm SE}$, which can be FM or antiferromagnetic (AFM).

In Fig.~\ref{JvsJU} we show the calculated $\Gamma^{i,i^\prime}_{\rm SE}$ for Cr$_8$ ($3d^3$, $S=3/2$) as a function of $U=U^{i,i}$ and $J=J^{i,i}$. 
The figure can be understood from the analytical expression of $\Gamma^{i,i^\prime}_{\rm SE}$
in the limit in which only density-density Coulomb interactions and leading order terms are retained,
\begin{equation}
\begin{split}
\Gamma^{i,i^\prime}_{\rm SE} \sim & \frac{2}{9} \sum^3_{n^\prime=1} \sum^5_{n=4} \frac{|t^{i,i^\prime}_{n^\prime,n}|^2+|t^{i,i^\prime}_{n,n^\prime}|^2}{U+\varepsilon_{n}-\varepsilon_n^\prime} \\+ & \frac{2}{9}\sum^3_{n^\prime=1}\sum^3_{n=1} \frac{|t^{i,i^\prime}_{n^\prime, n}|^2}{U+2J+\varepsilon_n-\varepsilon_{n^\prime}} \\-& \frac{2}{9}\sum^3_{n^\prime=1} \sum^5_{n=4} \frac{|t^{i,i^\prime}_{n^\prime, n}|^2+|t^{i,i^\prime}_{n,n^\prime}|^2}{U-3J+\varepsilon_n-\varepsilon_{n^\prime}} .
\end{split}\label{SE}
\end{equation}
Eq.~(\ref{SE}) shows the competition between the first two terms, which yield a positive, i.e., AFM contribution and the FM third term,  arising from excitations to empty states. For realistic parameters, $\Gamma^{i,i^\prime}_{\rm SE}$ is small and AFM.

The ZFS term $D^i$ in (\ref{HS}) {originates from the combined action of crystal-field and spin-orbit interactions. In the case of Cr$_8$ it is given by
\begin{eqnarray}
D^i 
&=& \frac{1}{2}\sum_{m} \frac{\left\langle \frac{3}{2},\pm \frac{3}{2} | \mathcal{\widehat{H}}_{SO} | m \right\rangle \left\langle m |\mathcal{\widehat{H}}_{SO}| \frac{3}{2},\pm \frac{3}{2} \right\rangle}{E_{\frac{3}{2} 
}-E_{m}} \\ \nonumber
&-&
\frac{1}{2}\sum_{m} \frac{\left\langle \frac{3}{2},\pm \frac{1}{2} | \mathcal{\widehat{H}}_{SO} | m \right\rangle \left\langle m |\mathcal{\widehat{H}}_{SO}| \frac{3}{2},\pm \frac{1}{2}\right\rangle}{E_{\frac{3}{2} 
}-E_{m}}.
\end{eqnarray}
Here  $\vert S,M \rangle$ {are many-electron states in  the $S=3/2$ ground multiplet with energy
 $E_{\frac{3}{2} 
 }$, while 
$|m \rangle$ {are all the excited multiplets connected to $\vert S,M \rangle$ by the spin-orbit interaction,
and have energy $E_{m}$.} 
Our calculations yield the full ZFS tensor, and thus we can identify the easy magnetization axis,
which in general is site-dependent. Remarkably, we find that
the molecular global $z$ axis (see above) is a nearly-easy axis for all sites.

\begin{table}[t]
   \centering
   \begin{tabular}{@{} l@{\;\;\;\;\;}rr@{\;\;\;\;\;}rr@{\;\;\;\;\;}rr @{\,}}
      &\multicolumn{2}{c}{Cr$_8$}&\multicolumn{2}{c}{Fe$_4$}&\multicolumn{2}{c}{Cr$_7$Ni} \\
      [0.5ex]
      \hline
      $\Gamma^{1,2}_{\rm SE}$ &1.99  &      &3.25&&2.10&\\
      $\Gamma_{\rm SE}^{2,3}$ &1.99  & & 0.15 & &1.99&
      \\ \\
            &Th &Exp $\!\!\sp{\tiny\mbox{ \cite{Cr8PRB}}}$ &Th &Exp $\!\!\sp{\tiny\mbox{\cite{Fe4INS}}}$&Th &Exp $\!\!\sp{\tiny\mbox{\cite{INSCr7Ni}}}$ \\[1ex]\hline
      $\Gamma^{1,2}$          &1.65  &1.46 &2.45&2.05 &1.75&1.70  \\
      $\Gamma^{2,3}$          &1.65  &1.46 &-0.08&-0.09 &1.65&1.46 \\
      $D^1$              &-0.06 &-0.03 &-0.03&&-0.48&-0.35 \\
      $D^2$              &-0.06 &-0.03 &-0.06&&-0.06&-0.03 \\
   \end{tabular}
   \caption{{ Top: Calculated super-exchange exchange couplings. Bottom: Calculated
   total magnetic couplings \cite{non-equiv} and zero-field splitting (Th) versus experiments (Exp).
   Sites $i=1,2,3$ are defined in Fig.~\ref{Cr8} and Fig.~\ref{Fe4}. In Cr$_7$Ni the Ni ion is on site 1 of Fig.~\ref{Cr8}.
  }}
   \label{exchange}
\end{table}

In the next paragraphs we discuss the results of our calculations (Tab.~\ref{exchange}),
in comparison with experiments. \\
Let us start from the AFR Cr$_8$.
Fig.~\ref{JvsJU} shows that,  for realistic $U$ and $J$ values, the super-exchange coupling $\Gamma^{i,i+1}_{\rm SE}$  is AFM and of the order of few meV. By using the value of $U$ and $J$ obtained in cLDA  (Tab.~\ref{exchange}) we obtain $\Gamma^{i,i+1}_{\rm SE}=1.99~$meV. We find that the ferromagnetic  direct Coulomb exchange is $\Gamma^{i,i+1}_{\rm CE}=-0.34~$meV. Hence, the total Heisenberg exchange constant is $\Gamma^{i,i+1}=1.65~$meV, AFM and in excellent agreement with experiments \cite{Cr8PRB}.   In addition, we find that the next-nearest neighbors exchange interaction is tiny {( $\Gamma^{i,i+2} \approx 10^{-2} ~ \Gamma^{i,i+1}$ )}; this explains why all experimental data can be interpreted on the basis of a nearest-neighbor spin Hamiltonian. Beside the dominant isotropic exchange coupling, we also find a sizable single-ion ZFS term. Our calculations yield a significant easy-axis anisotropy in the $z$ direction ($D^i<0$); non-axial terms are {an order of magnitude} smaller than $D^i$, in line with experiments. The calculated $D^i$ is twice the value extracted from inelastic neutron scattering data, a remarkably good agreement given the small value of $D^i$. \

Next we consider the SMM Fe$_4$ (Fig.~\ref{Fe4}). This molecule {has D$_3$ symmetry; three Fe$^{3+}$ ions ($3d^5$, $S=5/2$) are located at the vertices of an
equilateral triangle, and the fourth is at its center \cite{Fe4Chimica}. We find an AFM isotropic magnetic coupling between the central and external ions ($\Gamma^{1,2}$) and a small FM interaction between the external ions ($\Gamma^{2,3}$), { in excellent agreement with the values determined from experiments} \cite{Fe4INS}. We find that the super-exchange term is small for the external ions, and thus the FM Coulomb exchange dominates. Finally, we calculate the ZFS tensor and find again a nearly easy-axis anisotropy along $z$. {Our findings for $D^i$ differ of less than a factor 1.5 from the experimental results} \cite{notaFe4}.}

As last case we consider Cr$_7$Ni, an heteronuclear AFR that {can be obtained from Cr$_8$ by replacing a Cr$^{3+}$ ion with a Ni$^{2+}$ ($3d^8$, $S=1$).  This system is theoretically the most challenging, because two different types of ion (Cri$^{3+}$ and Ni$^{3+}$) are present. 
Again, we reproduce well all experimental results.
We find that the total Cr-Ni  isotropic coupling is AFM,  $\Gamma^{1,2} = 1.75~$meV, while the Cr-Cr
coupling $\Gamma^{2,2}$ is slightly smaller and close to the value obtained for Cr$_8$.
The ZFS parameters obtained for Ni$^{2+}$ ion \cite{spinorbitaNi} are again negative in sign (easy-axis anisotropy) and much larger than those of Cr$^{3+}$, in agreement with neutron spectroscopy results \cite{INSCr7Ni}.}

In conclusions, we present an {\em ab-initio} approach to calculate the terms of the spin Hamiltonians
for molecular nanomagnets. It is based on the construction of many-body Hubbard-like models, using Foster-Boys orbitals as a one-electron basis \cite{note}. 
We show that this scheme works remarkably well for MNMs.
For all systems considered, our results are closer to the experimental finding than those obtained by total-energy spin-configurations calculations based on the B3LYP functional \cite{b3lyp}.
Differently than spin-configurations  based approaches, our method allows us to determine the spin models without {\em a priori} assumptions on the form and the range of the Hamiltonian; furthermore, since it yields the parameters of the Hubbard model, it works also when charge fluctuations are sizable and the spin is not well defined, {like for molecules with metal-metal bonds} \cite{Longoni}, or when electrons are less localized,
such as in 4$d$ and 5$d$ systems, and 
can be used to calculate spectra. 
This scheme {is flexible, its complexity does not increase when the symmetry is low,
and it does not rely on the B3LYP or LDA+$U$ approximation to correlation effects.}
It could become essential for modeling MNMs whose spin Hamiltonian contain many anisotropic terms, in particular if the principal-axis directions and relative magnitude cannot be inferred simply by inspecting the molecular structure, as often the case for Co or $f$-electron systems. Finally, the many-body models for small MNMs can be still exactly solvable, allowing to test approximations often adopted but impossible to test in bulk correlated systems.
{Thus, we believe that our approach could become the method of choice for exploring fundamental issues and testing approximations, and for} identifying and designing new molecules for quantum devices.

Calculations were done on the J\"ulich supercomputer Juropa, grant number JIFF46.
E.P. acknowledges financial support from the Deutsche Forschungsgemeinschaft through research unit FOR1346.


\begin{thebibliography}{99}
\bibitem{explanations} All parameters $U^{i,i^\prime}_{m p m^\prime p^\prime}$ can be expressed as a function of the Slater integrals $F_0$, $F_2$ and $F_4$ \cite{book};
the average screened direct Coulomb integral for site $i$ is $U^{i,i}=F_0$ and the average screened exchange integral is defined as $J^{i,i}=(F_2+F_4)/14$.
\bibitem{non-equiv} In Cr$_8$ sites are only approximatively equivalent; the site variation of the exchange parameters are however small ($\sim4 \%$ for $\Gamma^{1,2}$ and $\sim 15 \%$ for $D^i$), and therefore we give the average values.
\bibitem{notaFe4} {Neutron spectroscopy data on the Fe$_4$ SMM have been interpreted by an effective Hamiltonian  for the entire molecule describing the splitting of the $S=5$ ground manifold. The dominant term is given by $D_{S=5} \left[ S_{z}^2-\frac{1}{3}S (S+1)\right]$, where $D_{S=5}=-55.8~\mu$eV. We have calculated $D$ using the microscopic $D^i$ parameter in Tab.~\ref{exchange} and find $D_{S=5}=-39~\mu$eV. It should be noted that in the calculation  of D we added also the (small) magnetostatic dipole-dipole interaction contribution evaluated in the point-dipole approximation. {It amounts to $-4.5 ~\mu$~eV}}.
\bibitem{note} A similar procedure, with localized Wannier functions instead of Foster-Boys orbitals, is used to determine magnetic and orbital couplings for strongly correlated transition-metal oxides \cite{book,eva}.
\end{thebibliography}
\end{document}